# Absence of ferromagnetism in V-implanted ZnO single crystals

Shengqiang Zhou[*], K. Potzger, H. Reuther, K. Kuepper, W. Skorupa, M. Helm, and J. Fassbender

*Institute for Ion Beam Physics and Materials Research, Forschungszentrum Rossendorf, POB 510119, 01314 Dresden, Germany*

---

[*] Electronic mail: s.zhou@fz-rossendorf.de

Abstract

The structural and magnetic properties of V doped ZnO are presented. V ions were introduced into hydrothermal ZnO single crystals by ion implantation with fluences of $1.2\times10^{16}$ to $6\times10^{16}$ cm$^{-2}$. Post-implantation annealing was performed in high vacuum from 823 K to 1023 K. The ZnO host material still partly remains in a crystalline state after irradiation, and is partly recovered by annealing. The V ions show a thermal mobility as revealed by depth profile Auger electron spectroscopy. Synchrotron radiation x-ray diffraction revealed no secondary phase formation which indicates the substitution of V onto Zn site. However in all samples no pronounced ferromagnetism was observed down to 5 K by a superconducting quantum interference device magnetometer.

I. Introduction

Diluted magnetic semiconductors (DMS) have recently attracted huge research attention because of their potential application for spintronics devices [1]. In DMS materials, transition or rare earth metal ions are substituted onto cation sites of the host semiconductor and are coupled with free carriers to yield ferromagnetism via indirect interaction [1, 2]. ZnO doped with V was found to be ferromagnetic at room temperature [3]. However, the origin of the observed ferromagnetism in transition metal doped ZnO is still controversial, e.g. ferromagnetic clusters [4], or extrinsic reasons (substrate impurities [5], use of stainless-steel tweezers [6]). The absence of ferromagnetism has been reported for Mn, Co doped ZnO [7]. Therefore, the observation of ferromagnetism in V doped ZnO should be carefully cross-checked. Ion implantation is a promising method to introduce magnetic metal ions into a host matrix at controlled concentrations and depths [8, 9]. In this work, V ions were introduced into semiconducting bulk ZnO by ion implantation. Rutherford backscattering/channeling (RBS/C) spectrometry, depth profiled Auger electron spectroscopy (AES), and synchrotron radiation x-ray diffraction (SR-XRD) are used to study the structural properties. A superconducting quantum interference device magnetometer (SQUID) is used to measure the magnetic response upon implantation and thermal annealing.

II. Experiments

Hydrothermally grown, commercially available ZnO single crystals, were implanted with V ions at 623 K and with different fluences from $1.2\times10^{16}$ to $6\times10^{16}$ cm$^{-2}$. The implantation energy of 300 keV yielded a projected range of $R_P=150\pm50$ nm (TRIM code). The post-implantation annealing was performed in a high vacuum ($p<10^{-7}$ mbar) furnace from 823 K to 1073 K for 15 minutes. The RBS/C spectra were collected with a collimated 1.7 MeV He$^+$ beam at a Van de Graaff accelerator with a surface barrier detector at 170°. SR-XRD was performed at the European Synchrotron Radiation Facility with monochromatic x-rays of 0.1541 nm wavelength. The virgin ZnO is pure diamagnetic with a susceptibility of -1.48×10$^{-6}$ emu/Oe·cm$^3$, and the background from ZnO has been subtracted in the following discussion. Depth profiling by AES was performed with a scanning Auger electron spectrometer Microlab 310F (Fisons Instruments).

III. Results and discussion

A. Lattice damage and recovering

RBS/C is used to check the lattice damage after implantation and recovering upon thermal annealing. The channeling spectra were collected by aligning the sample to make the impinging He$^+$ beam parallel with the ZnO<0001> axis. $\chi_{min}$ is the channeling minimum yield in RBS/C, which is the ratio of the backscattering yield at channeling condition to that for a random beam incidence [10]. Therefore, the $\chi_{min}$ labels the lattice disordering degree upon implantation, and an amorphous sample shows a $\chi_{min}$ of 100 %, while a perfect single crystal corresponds to a $\chi_{min}$ of 1-2 %. Figure 1 shows the representative RBS/C spectra for different fluences and annealing temperatures. The arrow labeled Zn indicates the energy for backscattering from surface Zn atoms. The humps in the channeling spectra mainly result from the lattice disordering due to implantation [11]. The RBS/C measurements revealed that the ZnO host material partly remains crystalline after V-implantation from 1.2×10$^{16}$ up to a fluence of 6×10$^{16}$ cm$^{-2}$. $\chi_{min}$ is increased from 57% to 79% with increasing fluence. Thermal annealing can partially recover the damaged lattice, and the $\chi_{min}$ is decreased from 57% to 35% with increasing annealing temperature for the low concentration sample.

B. Thermal mobility of the V ions

Depth profiled AES was performed to check the mobility of V with thermal treatment. Figure 2 shows the result for the sample with a fluence of $6\times10^{16}$ cm$^{-2}$. The sputtering time quantitively represents the depth. With annealing at 823 K, V ions start to diffuse outside of ZnO, and after annealing at 1073 K, the V profile vanishes almost completely. Thus, V ions become rather mobile inside ZnO with thermal annealing.

C. Excluding crystalline secondary phase

Figure 3(a) shows the SR-XRD patterns for the sample with a fluence of $6\times10^{16}$ cm$^{-2}$ in a Bragg-Brentano geometry. No evident crystalline V, or V-oxide nanoparticles could be detected. In order to confirm this conclusion, grazing incidence measurements were performed. GIXRD has been successfully used to detect Mn-silicides nanoparticles in Mn-implanted Si [12]. Figure 3(b) shows the GIXRD patterns for the same sample as in Figure 3(a). Again, there is no secondary phase detectable. Therefore it is assumed that the majority of implanted V ions are substituted onto Zn sites. Note the asymmetry of the diffraction peaks in Figure 3(a), shoulders on the right side (smaller lattice constant) are clearly observed. These shoulders decrease with increasing annealing temperature, and can therefore be associated with lattice damage or ZnO substituted with V since V has a smaller volume than Zn. In view of a detailed study on ion implantation into GaN where the implantation induces a lattice expansion of GaN (a shoulder at left side) [13], we would rather attribute the observed shoulders to ZnO substituted with V. With annealing, V ions are more mobile and the substitution is reduced, which has been shown by AES measurement.

D. Magnetic properties

Figure 4(a) shows the magnetization versus field curves (M-H loops) recorded at 5 K for all samples. Obviously none of the V-implanted ZnO samples show any pronounced ferromagnetism. Figure 4(b) shows the temperature dependent magnetic susceptibility of the samples with a high fluence. All samples exhibit mainly the diamagnetic background from the substrate with a very weak paramagnetic component. The paramagnetic component shows Curie-Weiss behaviour. However, the M-H curve for the as-implanted sample with a fluence of $6\times10^{16}$ cm$^{-2}$ almost looks like a 'noisy' hysterisis loop. In order to confirm if there is a weak ferromagnetic contribution in this sample, temperature dependence magnetization was additionally measured under both zero field cooled (ZFC) and field cooled (FC) conditions. The difference between FC and ZFC magnetization is a measure of the ferromagnetic contribution [8]. Figure 4(c) shows a slight difference (below $10^{-9}$ emu/mg) between ZFC/FC magnetization, which confirms the likelihood of a weak ferromagnetic contribution. Therefore, the application of a high V fluence under low processing temperatures could be an approach to enhance the ferromagnetic properties.

IV. Conclusions

V-implanted ZnO crystals were studied to correlate their structural and magnetic properties. ZnO is a very hard material under ion bombardment, and the lattice damage can be partially recovered by thermal annealing. V ions are thermally mobile and diffuse into the ZnO bulk upon thermal annealing. Upon implantation and annealing, no crystalline secondary phase can be detected by SR-XRD. However, all the samples, with different V concentration and thermal processing, show no pronounced ferromagnetism. The latter effect does not necessarily originate from the general lack of indirect magnetic coupling in such systems, but rather from the low initial charge carrier concentration of the hydrothermal samples (usually from $10^9$ to $10^{12}$ cm$^{-3}$) that is not sufficiently increased by the defects resulted from implantation.

Figure Captions

Fig. 1 (Color online) Representative RBS/channeling spectra with different fluence and thermal processing. The arrow on the right side gives a sequence of all spectra. The top one is a random spectrum, while others are channeling ones.

Fig. 2. (Color online) AES depth profiles of V of the high fluence ($6\times10^{16}$ cm$^{-2}$) sample in the as-implanted state and after annealing.

Fig. 3 (Color online) (a) SR-XRD symmetric $2\theta/\omega$ scan, (b) grazing incidence scan for the same series of samples.

Fig. 4 (Color online) (a) Hysteresis loops measured at 5 K for V doped ZnO. Numbers of $1.2\times10^{16}$ and $6\times10^{16}$ label the V fluence at cm$^{-2}$, and process temperatures are indicated. (b) M(T) curves for samples with the fluence of $6\times10^{16}$ cm$^{-2}$. Inset shows the same measurement with a smaller scale, revealing no clear difference in the as-implanted and annealed samples. (c) ZFC/FC magnetization curves for the as-implanted sample with the fluence of $6\times10^{16}$ cm$^{-2}$.

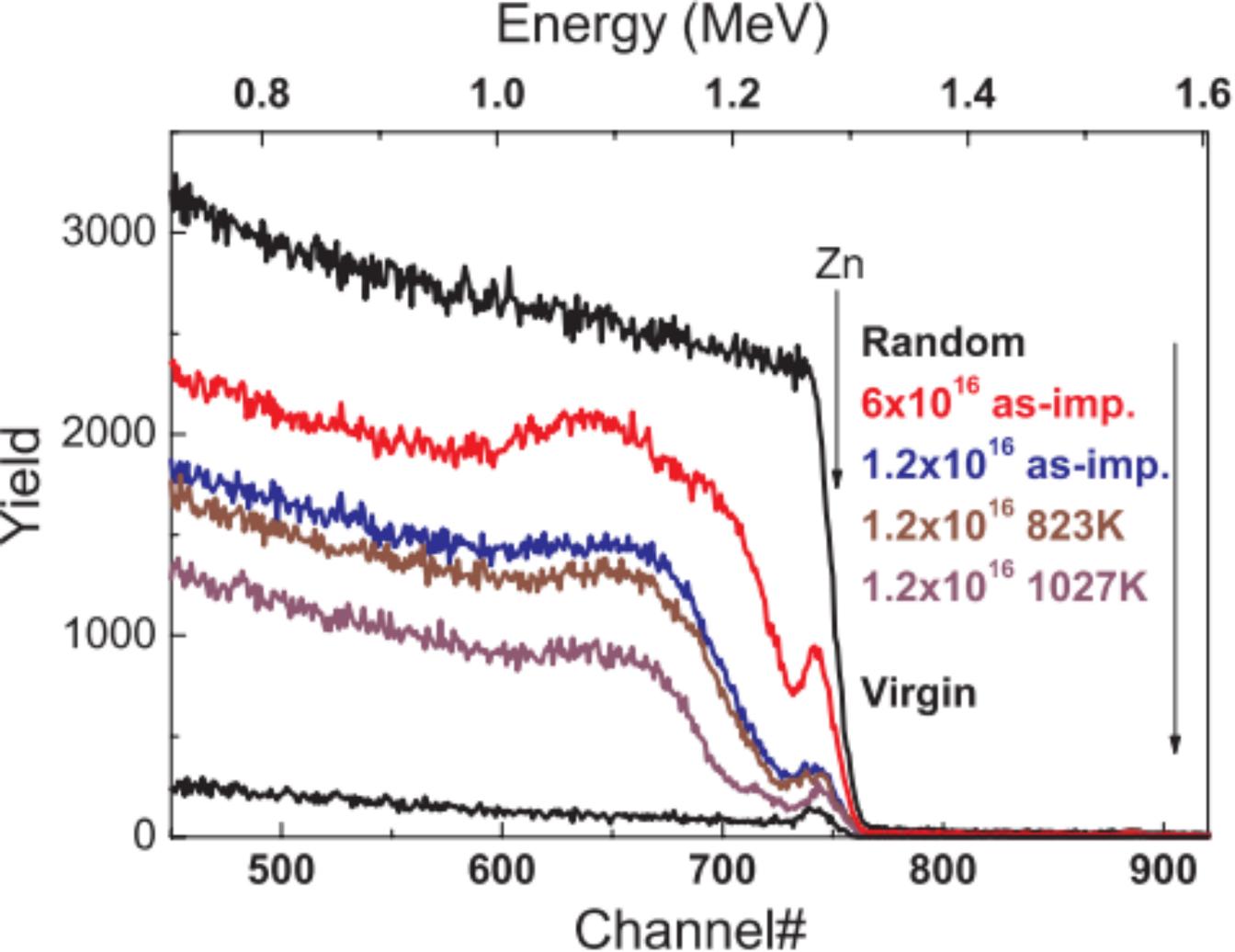

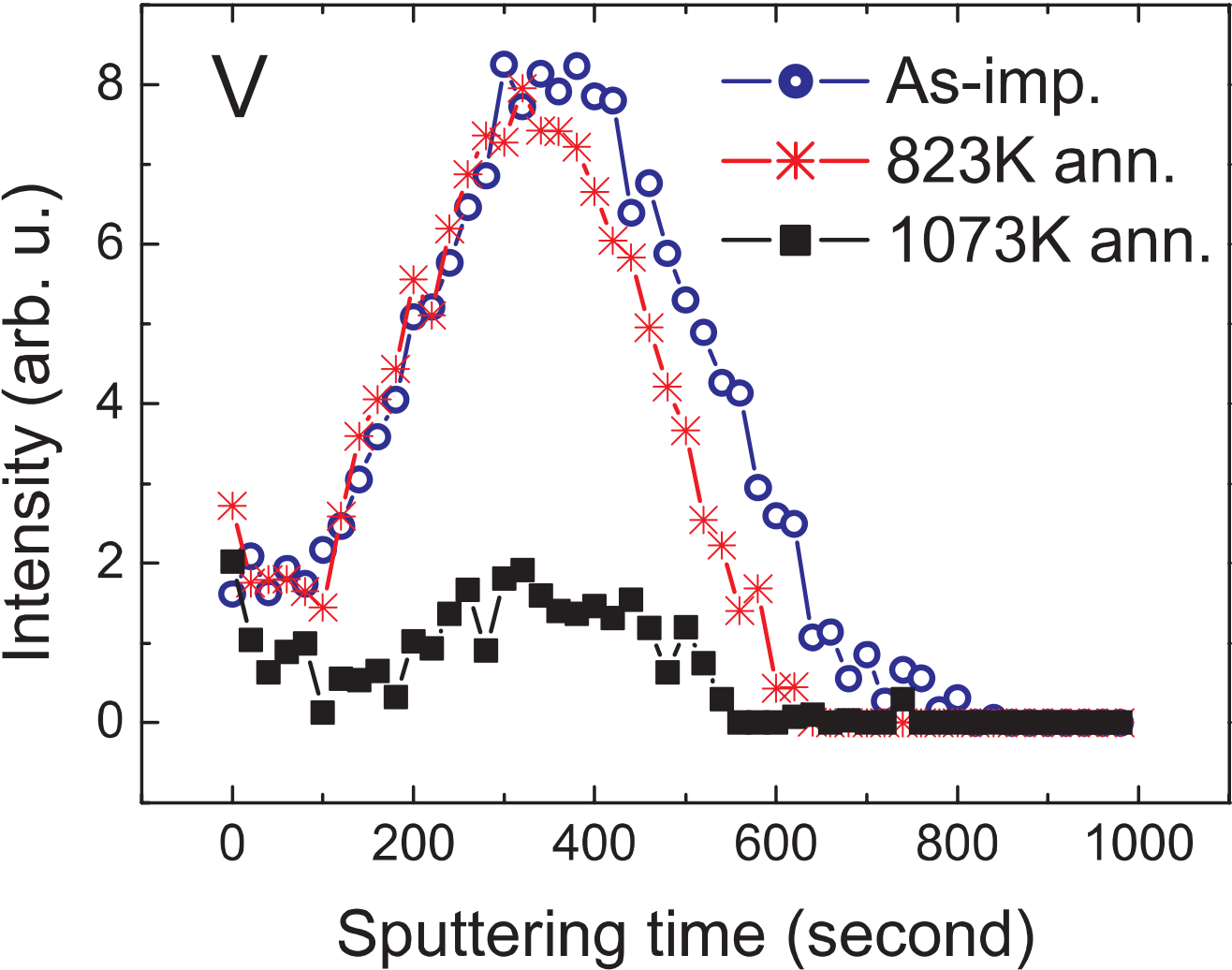

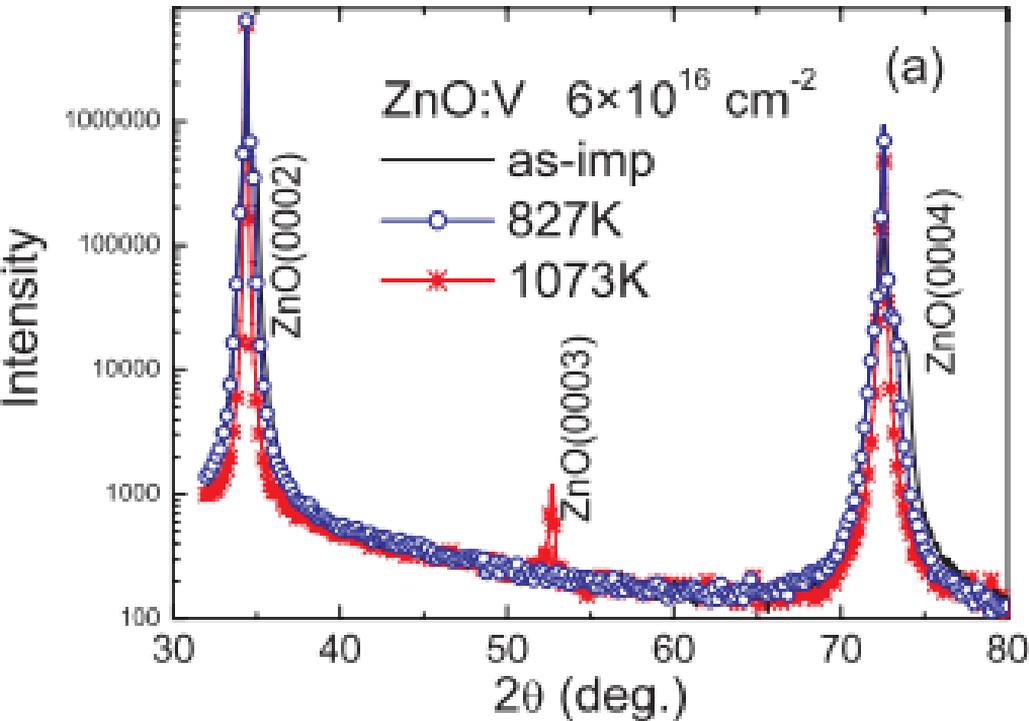
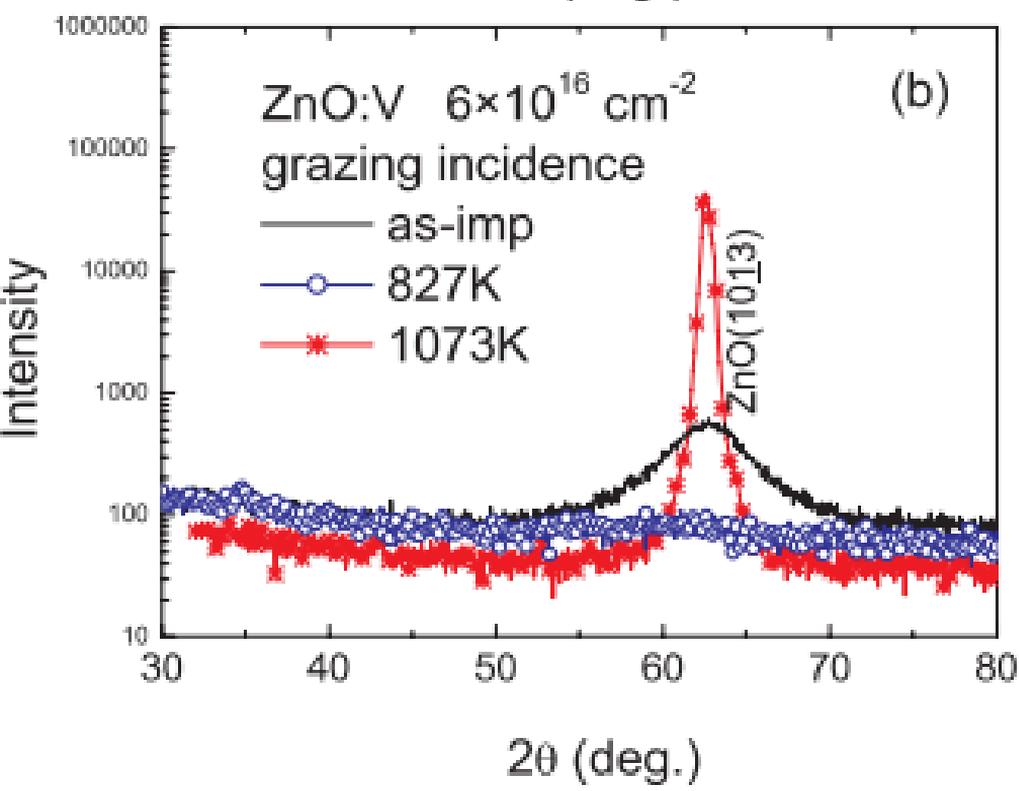

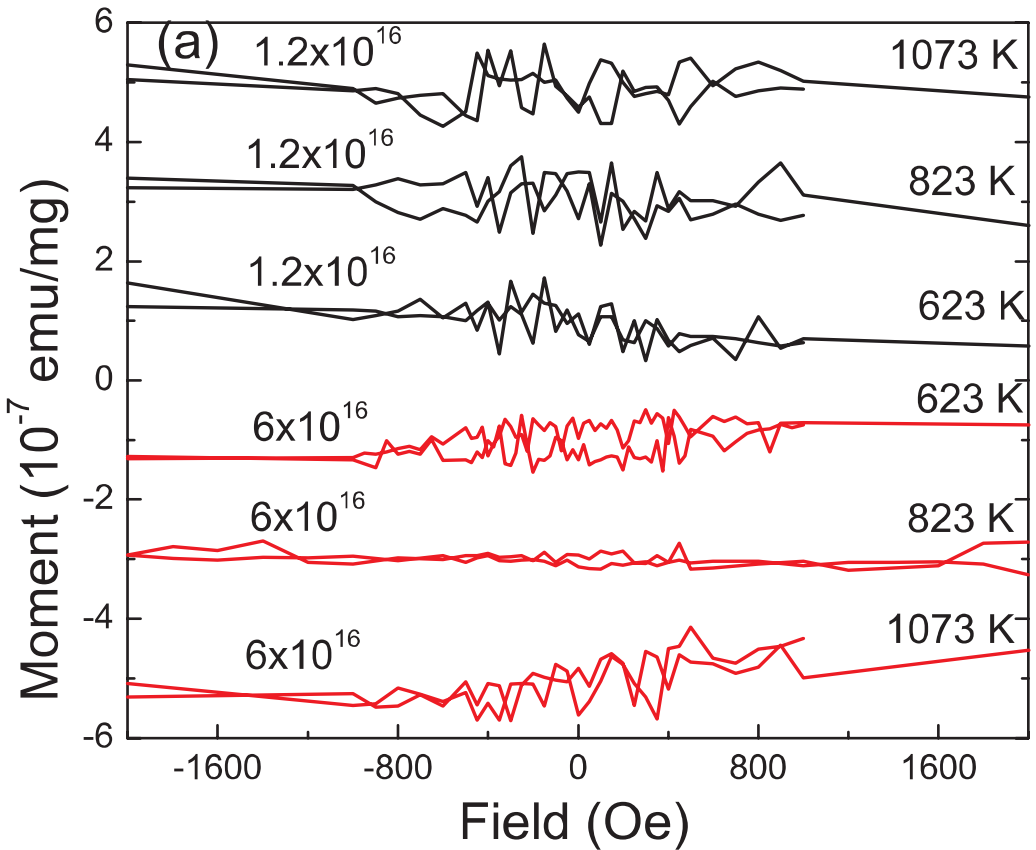
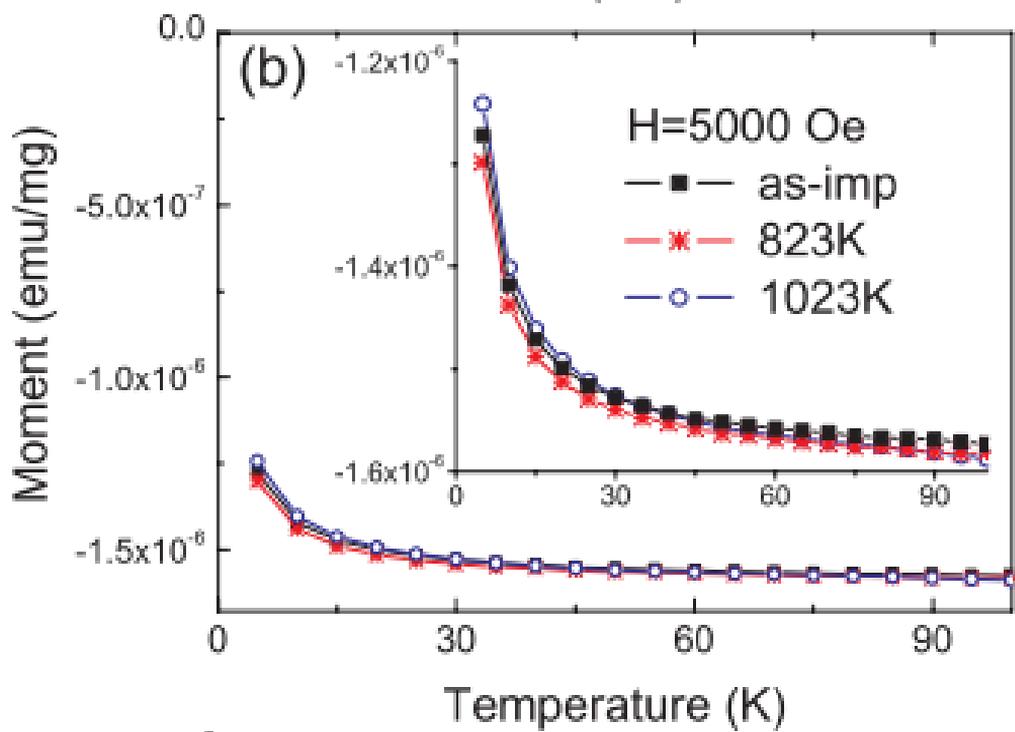
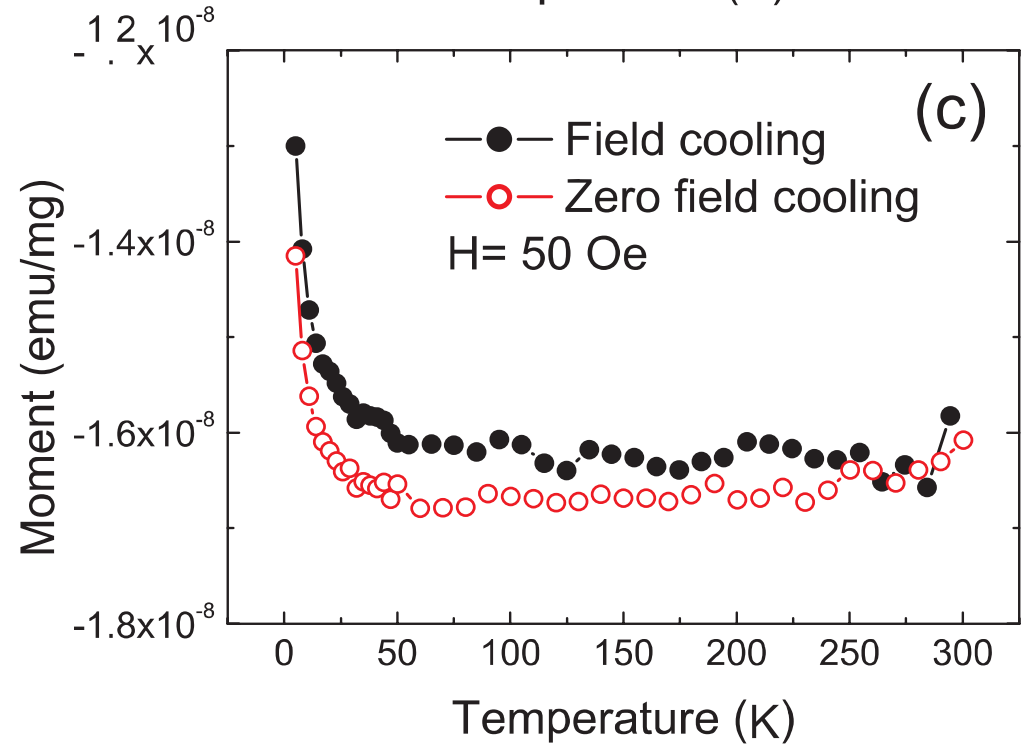